\documentclass[aps,prb,reprint,superscriptaddress,amsmath,amssymb]{revtex4-1}
\usepackage{epsfig,psfrag}
\usepackage{dcolumn}
\usepackage{bm}
\usepackage{graphicx}
\usepackage{color}

\begin{document}
\title{Disorder effects in InAs/GaSb topological insulator candidates}

\author{C. Ndebeka-Bandou}
\affiliation{Institute for Quantum Electronics, ETH Zurich, Auguste-Piccard-Hof 1, 8093 Zurich, Switzerland} 
\author{J. Faist}
\affiliation{Institute for Quantum Electronics, ETH Zurich, Auguste-Piccard-Hof 1, 8093 Zurich, Switzerland}

\begin{abstract}
We report the theoretical investigation of the disorder effects on the bulk states of inverted InAs/GaSb quantum wells. As disorder sources we consider the interface roughness and donors/acceptors supplied by intentional doping. We use a $\bm{k}\cdot\bm{p}$ approach combined with a numerical diagonalization of the disordered Hamiltonian to get a full insight of the disordered eigenenergies and eigenfunctions of the electronic system.
While interface roughness slightly pertubs the carrier motion, we show that dopants strongly bind and localize the bulk states of the structure. Moreover, both types of scatterers strengthen the intrinsic hybridization between holes and electrons in the structure.
\end{abstract}


\maketitle

\section{Introduction}
\label{sec1}

Topological insulators are phases of matter characterized by an insulating gap and protected conducting surface (or edge) states \cite{hasan2010}. In two-dimensional (2D) structures, these phases manifest themselves by the formation of helical channels at the edges of an insulating 2D electronic system, giving rise to the so-called Quantum Spin Hall Effect (QSHE). After its theoretical prediction \cite{bernevig2006,bernevig_science2006} and its first experimental demonstration in HgTe quantum wells (QWs) \cite{konig2007}, it has been predicted that InAs/GaSb QWs should also be promising 2D topological insulator candidates \cite{liu2008}.
Similarly to HgTe QWs, InAs/GaSb QWs have unique band alignments that display a subband inversion transition as a function of the layer thickness. Moreover, due to their type II alignment (where the confinement of holes and electrons occurs in two different layers), InAs/GaSb QWs have the additional advantage of being electrically tunable through the phase transition \cite{yang1997,cooper1998}.

A non-local transport through edge states has been already reported in inverted InAs/GaSb (QWs) \cite{knez2011,nichele2014,mueller2015}. However, a substantial bulk conductivity still degrades the visibility of the dissipationless edge channels indicating that the bulk is rather a  metallic system. A few solutions have so far been proposed such as the intentional addition of disorder in the sample to lower the bulk mobility and suppress the parasitic channels \cite{suzuki2013,charpentier2013,du2015}. 
In contrast to experimental studies, there has been little theoretical work done on disordered inverted InAs/GaSb QWs. A theoretical study of QSH states in doped InAs/GaSb QWs has been reported by Xu \textit{et al.} \cite{xu2014}, showing that an intentional Si-doping leads to the opening of a mobility gap. In this work, the dopants were considered as delta-like scatterers within the tight-binding approach, not accounting for the actual long-range of the Coulomb interactions. This approach is sufficient to predict the existence of in-gap localized states but is not adequate to get a quantitative description of the disordered bulk states in these complex structures.

In this paper, we present a theoretical analysis of the disorder effects on the bulk states in inverted InAs/GaSb QWs based on  realistic disorder modellings. We combine an eight-band $\bm{k}\cdot\bm{p}$ model and a numerical diagonalization of the disordered Hamiltonian within the envelope function formalism. This approach is more appropriate for the description of slowly varying potentials like Coulomb potentials and allows an accurate determination of the band dispersion of the non-disordered structure as well as a full treatment of the carrier-disorder interactions. 

Our paper is organized as follows. In Sec.~\ref{sec2}, we first present our model of disorder where we concentrate on two types of static scatterers: the interface roughness and the donors/acceptors supplied by intentional doping.  Then, in Sec.~\ref{sec3} we describe our theoretical approach based on the $\bm{k}\cdot\bm{p}$ model for the computation of the unperturbed states of a AlAs/InAs/GaSb/AlAs heterostructure and the numerical diagonalization of the disordered Hamiltonian expanded in this unperturbed basis. Finally, in Secs.~\ref{sec4} and \ref{sec5}, we report the different disorder effects such as the binding effects and the spatial localization of the states as well as the disorder-induced hybridization between valence and conduction carriers. 

\section{Model of disorder}
\label{sec2}

\begin{figure}
 \includegraphics{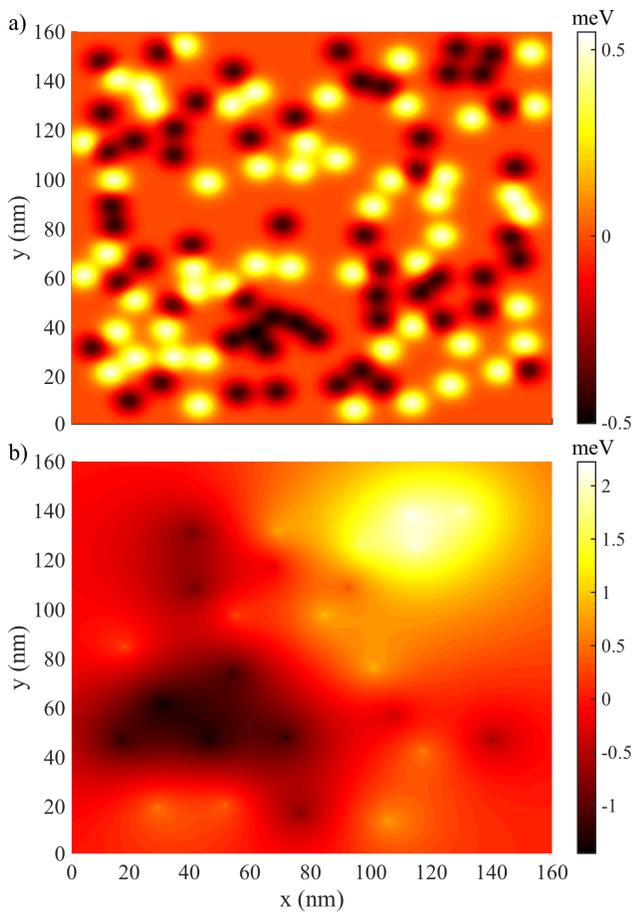}
 \caption{(Color online) Potential energy created by a) interface defects and b) Si-dopants in a 50/15/8/50~nm AlAs/InAs/GaSb/AlAs structure. The scatterers have randomly generated in-plane coordinates and their location along the growth axis coincides with the InAs/GaSb interface. The sample size is $S$ is 160$\times$160~nm$´^{2}$ and the disorder parameters are given in the text.}
\label{fig1}
\end{figure}

The interface roughness results from the interdiffusion of the layers during the growth process. Therefore it is an unavoidable source of scattering in heterostructures. In the envelope function formalism, the interface defects can be modeled by randomly distributed Gaussian protusions either from the InAs layer in the GaSb one (attractive defects) or vice versa (repulsive defects) \cite{ndebeka2012}. Using this model and considering a single interface located at $z=z_0$ ($z$ defines the growth axis), we write the disorder potential as:
\begin{equation}
 V_\mathrm{def}(\bm{\rho},z)=V_b\sum_{j=1}^{N_\mathrm{def}}f_j(z) \exp\left(-\frac{(\bm{\rho}-\bm{\rho}_j)^2}{2\sigma^2}\right),
 \label{eq_Vdef}
 \end{equation}
where $f_j(z)=Y(z-z_0)Y(h-z+z_0)$ ($f_j(z)=-Y(-z+z_0)Y(h+z-z_0)$) for repulsive (attractive) defects and with $Y$ the Heaviside function. The roughness height $h$ is set to two monolayers. The in-plane coordinates $\bm{\rho}_j=(x_j,y_j)$ are random and locate the $j^\mathrm{th}$ protusion on the surface $S$ of the sample. The in-plane extension of the defects $\sigma$ and the band offset discontinuity $V_b$ are set to $5.6$~nm and 150~meV respectively while the number of defects $N_\mathrm{def}$ is fixed by the fractional coverage of the surface $f=\pi\sigma^2N_\mathrm{def}/S$. In our model, the numbers of attractive and repulsive defects are equal. Figure~\ref{fig1}a shows the potential energy created by a random distribution of interface defects in a AlAs/InAs/GaSb/AlAs quantum well.

Due to their ability to bind and localize states \cite{ndebeka2013}, dopants are also known to be a relevant source of scattering in heterostructures. We consider the doping configuration that has been experimentally realized by Du \textit{et al.} \cite{du2015} where one monolayer of Si atoms of concentration $n_\mathrm{imp}=10^{11}$~cm$^{-2}$ is grown at the InAs/GaSb interface. 
Since Si atoms behave as donors in InAs and as acceptors in GaSb \cite{shen1993}, the densities of donors and acceptors are both equal to $n_\mathrm{imp}/2$ in our model. Like interface defects, the dopants are randomly distributed in the layer plane located at the InAs/GaSb interface. Accounting for screening effects by the mobile carriers, the disorder potential is described by a Yukawa-like potential with Debye screening length $\lambda$ \cite{ndebeka2016}:
\begin{equation}
 V_\mathrm{dop}(\bm{r})=\sum_{j=1}^{N_\mathrm{imp}}s_j\frac{e_0^2}{4\pi\varepsilon_0\varepsilon_r}\frac{ e^{-|\bm{r}-\bm{r}_j|/\lambda}}{|\bm{r}-\bm{r}_j|},
 \label{eq_Vimp}
 \end{equation}
where $\bm{r}_j=(\bm{\rho}_j,z_0)$, $e_0$ is the elementary charge, $\varepsilon_0$ and $\varepsilon_r$ are the dielectric constants of the vacuum and the material respectively. The prefactor $s_j$ equals $+1$ ($-1$) for acceptors (donors).  Figure~\ref{fig1}b shows the potential energy created by a random distribution of coulombic donors and acceptors in the same structure as in Fig.~\ref{fig1}a.

\section{$\bm{k}\cdot \bm{p}$ model and numerical diagonalization}
\label{sec3}

As previously, we consider a 15~nm/8~nm InAs/GaSb QW embedded in two wide gap AlSb barriers of thickness 50~nm. In the presence of disorder, the one-particle envelope Hamiltonian is
\begin{equation}
 H=H_0+V,
 \label{eq_hamiltonian}§		
\end{equation}
where $H_0$ is the Hamiltonian in absence of disorder and $V$ is the disorder potential either equal to $V_\mathrm{def}$ or $V_\mathrm{dop}$.  We compute the eigenstates of $H_0$, i.e. the subband dispersion $E_n(\bm{k})$ and the spinor wavefunctions $\bm{\chi}_n(\bm{k},z)$, for all $\bm{k}$ directions by using the eight-band $\bm{k} \cdot \bm{p}$ model developped in Ref.~\onlinecite{zakharova2001}. $n$ labels the different subbands and $\bm{k}=(k_x,k_y)$ is the in-plane momentum.  The direction (001) coincides with the growth axis and the material parameters are taken from Ref.~\onlinecite{halvorsen2000}. 
 
\begin{figure}
 \includegraphics{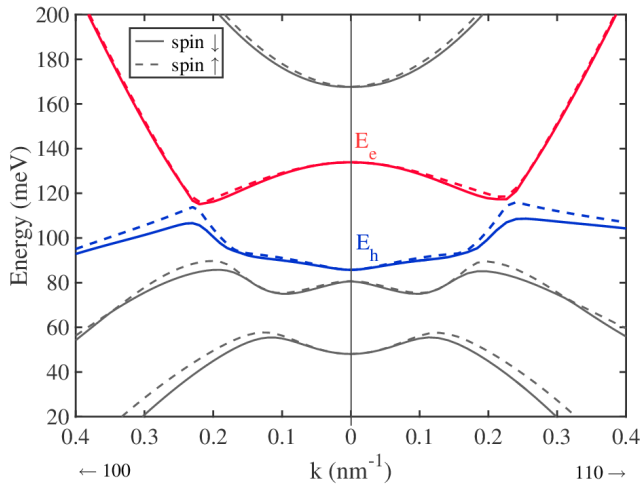}
 \caption{(Color online) Subband dispersion along the (100) (left panel) and (110) (right panel) directions of a 50/15/8/50~nm AlAs/InAs/GaSb/AlAs structure. The spin-down dispersion is plotted in solid line and the spin-up dispersion in dashed line.  }
\label{fig2}
\end{figure}

Figure~\ref{fig2} shows the calculated subband dispersion of the InAs/GaSb QW along the (100) and (110) directions for the two spin orientations. As expected from the layer ratio, the structure exhibits a band-inversion characterized by multiple anti-crossings at finite $k$ values, along with the opening of a hybridization gap between the conduction subband $E_e$ and the valence one $E_h$. Moreover, due to the asymmetry of the confining potential, the spin degeneracy is lifted at finite $k$. This results in a significant dependence of the dispersion upon the spin orientation \cite{silva1997}. The spin-up dispersion has a gap of about 6~meV while the spin-down dispersion is nearly gapless (an energy gap of less than 1~meV is extracted from the data). Since such a small gap is not wide enough to prevent parasitic scatterings between the conduction and valence bulk states, this local closing of the energy gap can be readily identified as the main cause of the metallic behaviour of the structure observed experimentally.  

In the following, we consider the case where the Fermi level lays inside the hybridization gap. Thus, for the numerical diagonalization of $H$, we restrict our consideration to the two subbands of interest: $E_e$ and $E_h$ (see Fig.~\ref{fig2}). Since $V$ is spin-independent, the results obtained for a given spin orientation are qualitatively similar to those obtained for the opposite spin orientation, the difference between the two configurations mainly resides in the size of the energy gap.
The calculation of the eigenstates of (\ref{eq_hamiltonian}) is performed by a numerical diagonalization of $H$ in the two-subband basis $\{E_e,E_h\}$ of periodic plane waves (Born-von Karman boundary conditions) on a $160\times160$~nm$^2$ finite sample \cite{ndebeka2012}. The wavefunctions of the basis states are 
\begin{equation}
 \Psi(\bm{\rho},z)=\langle \bm{\rho},z|n,\bm{k}\rangle=\frac{1}{\sqrt{S}}e^{i\bm{k}\cdot\bm{r}}\bm{\chi}_n(\bm{k},z) \ ,
\end{equation}
where $n=\{\mathrm{e,h}\}$. For a basis formed by $N$ periodic plane waves, the Hamiltonian $H$ is expressed as the following $2N\times2N$ block matrix 
\begin{equation}
 H=\begin{pmatrix}
    H_\mathrm{ee} &  H_\mathrm{eh}\\
    H_\mathrm{he} & H_\mathrm{hh}
   \end{pmatrix}.
   \label{matrixH}
\end{equation}
The diagonal blocks in (\ref{matrixH}) contain the intra-subband coupling terms:
%
\begin{multline}
 H_{nn}=\\\begin{pmatrix}
    E_n(\bm{k}_1)+V_{\bm{k}_1,\bm{k}_1}^{nn} & V_{\bm{k}_1,\bm{k}_2}^{nn} & \cdots&  V_{\bm{k}_1,\bm{k}_N}^{nn}\\
    & & & \\
    V_{\bm{k}_2,\bm{k}_1}^{nn} & \ddots & & \vdots \\
    \vdots & & \ddots & \vdots\\
    \vdots & & & V_{\bm{k}_{N-1},\bm{k}_N}^{nn} \\
    & & & \\
    V_{\bm{k}_{N},\bm{k}_1}^{nn}  & \cdots &  V_{\bm{k}_{N},\bm{k}_{N-1}}^{nn} & E_n(\bm{k}_N)+V_{\bm{k}_N,\bm{k}_N}^{nn}
   \end{pmatrix},
   \label{matrixH_diagonal}
\end{multline}
%
while the off-diagonal blocks contain the inter-subband coupling terms
\begin{equation}
 H_{nn'}=\begin{pmatrix}
    V_{\bm{k}_1,\bm{k}_1}^{nn'} & \cdots & V_{\bm{k}_1,\bm{k}_N}^{nn'}\\
    \vdots & \ddots & \vdots \\
 V_{\bm{k}_N,\bm{k}_1}^{nn'} & \cdots & V_{\bm{k}_N,\bm{k}_N}^{nn'}
   \end{pmatrix}.
   \label{matrixH_off_diagonal}
\end{equation}
In (\ref{matrixH_diagonal}) and (\ref{matrixH_off_diagonal}), the matrix elements of the disorder potential $V_{\bm{k},\bm{k'}}^{nn'}=\langle n\bm{k} | V|n'\bm{k'}\rangle$ are calculated the same way as in Ref.~\onlinecite{ndebeka2014}.

The disorder breaks the in-plane translational invariance, giving rise to a set of discrete energy levels for the energy spectrum. Consequently, the in-plane momentum is no longer a good quantum number and must be replaced by a discrete index $\nu$, namely:
\begin{equation}
 \bm{k}\rightarrow \nu \quad;\quad E_{n}(\bm{k})\rightarrow E_\nu,
\end{equation}
where the $E_\nu$'s are the eigenvalues of $H$. Their corresponding eigenfunctions are then written as
\begin{equation}
 \Psi_\nu(\bm{\rho},z)=\frac{1}{\sqrt{S}}\sum_{\bm{k}}\sum_{n=e,h}c_n^{(\nu)}(\bm{k})e^{i\bm{k}\cdot \bm{\rho}} \bm{\chi}_n(\bm{k},z),
 \label{eq_eigenstate}
\end{equation}
where the coefficients $c^{(\nu)}_n(\bm{k})$ are the coordinates of the eigenvectors of $H$ associated with the eigenvalue $E_\nu$.
This approach consists of a full treatment of the static disorder where the only assumption lies in the choice of the expression and the parameters of $V$.
Due to the randomness of the disorder potential, the diagonalization of $H$ is repeated for a large number of random realizations. The quantities that are showed in the following are the results of an averaging over a large set of random trials. 

\section{Binding and localization effects}
\label{sec4}

\begin{figure}
 \includegraphics{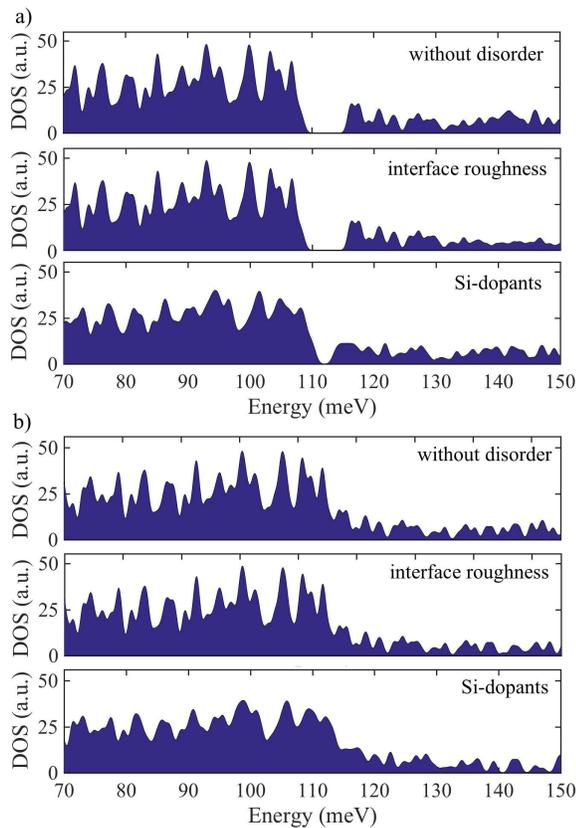}
 \caption{(Color online) Density of states of a 50/15/8/50~nm AlAs/InAs/GaSb/AlAs structure for a) spin-down states and b) spin-up states. Three cases are considered: in the absence of disorder (upper panel), in the presence of interface defects (middle panel) and Si-dopants (lower panel). 50 random realizations of disorder have been averaged.}
\label{fig3}
\end{figure}

As a first noticeable effect, the disorder reorganizes the energy spectrum of the system. Figure~\ref{fig3} displays the density of states (DOS) in three different configurations: without disorder, with interface roughness and with Si-dopants. The results for the spin-down and spin-up orientations are shown in Fig.~\ref{fig3}a and \ref{fig3}b respectively. Both types of scatterers create states below (for conduction states due to attractive potentials) and above (for valence states due to repulsive potentials) the subband edges. The effect of interface roughness is quantitatively much smaller. 
Extremely shallow bound states with typical binding energies of 0.3-0.5~meV below/above the subband edges are found whereas binding energies of dopants reach 2-3~meV and tend to close the energy gap. 
This very weak binding effect by the interface defects originates from the shorter-range and smaller-depth compared to the fluctuations of the Coulomb potential. Note also that in type II QWs, the carriers are mostly confined in each layer and therefore have a weak probability density $|\bm{\chi}_n(\bm{k},z_0)|^2$ at the interface \cite{zakharova2001}, where both interface defects and dopants  are actually located. This reduces the potential strength of both scatterers \cite{ndebeka2013apex}.
Similarly, bound states are also found for the spin-up orientation. However, a plot of the DOS cannot reveal their existence due to the quasi-absence of energy gap. In Fig.~\ref{fig3}b one can still notice a broadening of the DOS arising from the lifting of degeneracy of the states by the dopants.

\begin{figure}
 \includegraphics{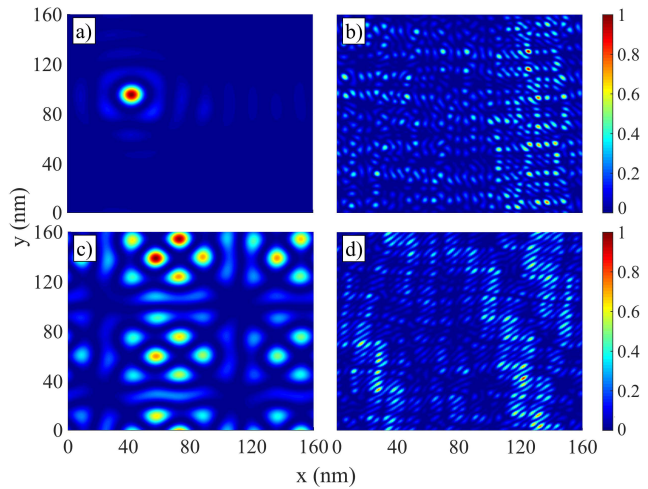}
 \caption{(Color online)  In-plane probability density of disordered states at various energies in a 50/15/8/50~nm AlAs/InAs/GaSb/AlAs structure. The upper panels correspond to the case where dopants are take as source of disorder in the structure and the energies of the states are a) $E_\nu$=114~meV (in-gap state) and b) $E_\nu$=145~meV (continuum state). The lower panels correspond to the case where interface roughness is taken as source of disorder and the energies of the states are c) $E_\nu$=115.6~meV (in-gap state) and d) $E_\nu$=141.5~meV (continuum state). The two plots of each panel correspond to the same random realization of disorder.}
\label{fig4}
\end{figure}

Another way to evidence bound states is to evaluate their spatial localization. Bound states emerge from multi-scattering events and have the particularity to be spatially localized in the layer plane due to the admixture of the carrier in-plane wavefunctions by the disorder \cite{ndebeka2013}. Fig.~\ref{fig4} displays the in-plane probability density of disordered states at various energies in the presence of dopants (upper panels) and interface defects (lower panels). In-gap states are strongly localized by the dopants, as shown in Fig.~\ref{fig4}a, whereas continuum states remain extended (see Fig.~\ref{fig4}b). As shown in Fig.~\ref{fig4}c, the localization by interface roughness of shalow bound states is weak. The continuum states in both disorder configurations have a similar extension as shown in Figs.~\ref{fig4}b and \ref{fig4}b.

One way to quantify this localization effect is to compute the in-plane localization length of the disordered states. The in-plane localization can be defined as
\begin{equation}
 l_\nu=\left(\iint d^2\bm{\rho}  |\varphi_\nu (\bm{\rho})|^4 \right)^{-1/2},
\end{equation}
where $\varphi_\nu (\vec{\rho})$ is the in-plane part of the wavefunction obtained by integrating the modulus square of  (\ref{eq_eigenstate}) over $z$.

\begin{figure}
 \includegraphics{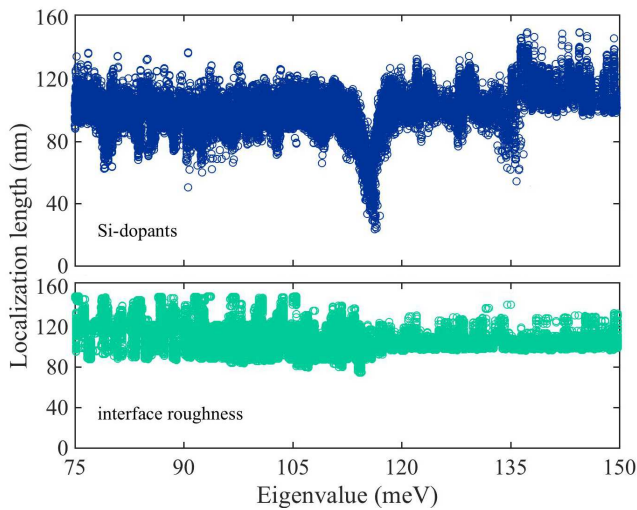}
 \caption{(Color online) In-plane localization length versus the eigenvalues of the bulk states of a 50/15/8/50~nm AlAs/InAs/GaSb/AlAs structure. Two type of scatterers are considered: the Si-dopants (upper panel) and the interface roughness (lower panel). These results are shown for the spin-up orientation. 50 random realizations of disorder have been taken into account.  }
\label{fig5}
\end{figure}

Figure~\ref{fig5} shows the localization length of the spin-up eigenstates. The spatial localization by the dopants is significantly more pronounced, especially close to the subband edges (around 116~meV, see also Fig.~\ref{fig1}) where a mobility gap of about 4~meV opens. On the other hand, the localization by interface defects remains weak as expected from the in-plane probability density computed in Figs.~\ref{fig4}c and \ref{fig4}d. Similar features are found for the spin-down states.  

These results are consistent with the experimental observations \cite{du2015} and the results found by Xu \textit{et al.} with the tight-binding formalism and additionally provide a quantitative estimate of the binding energies and localization lengths. In particular, a localization of the continuum states by dopants is found to be about 25\% compared to non-disordered extended states on a 160$\times$160 nm$^2$ surface \cite{footnote}. An intentional doping certainly closes the energy gap and decreases the bulk conductivity by opening a mobility gap but also disturbs the remaining extended states.

\section{Disorder-induced hybridization}
\label{sec5}

\begin{figure}
 \includegraphics[scale=1]{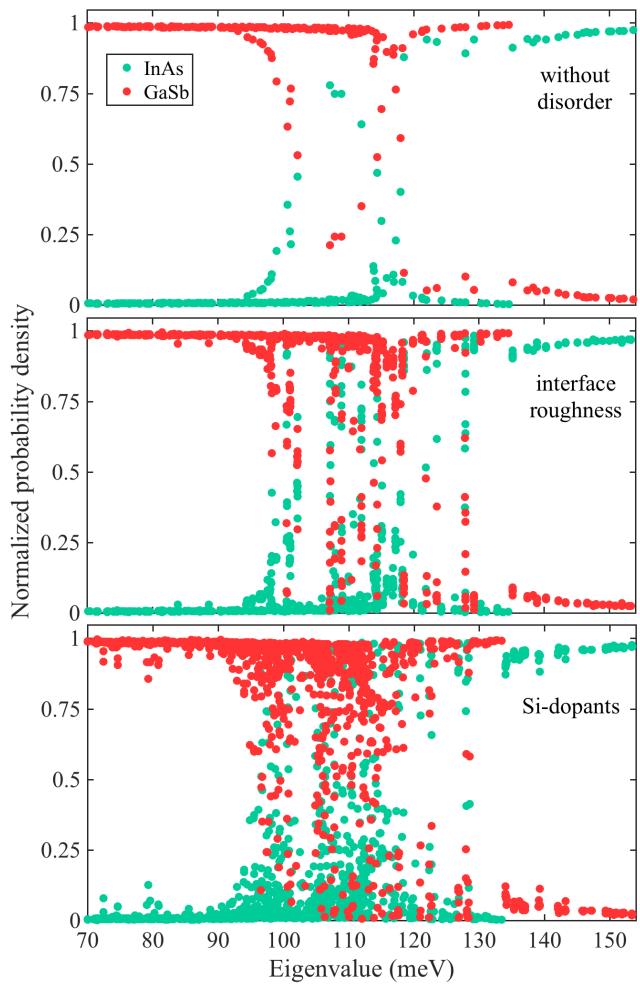}
 \caption{(Color online) Normalized probability density in each layer of the spin-up states in a 50/15/8/50~nm AlAs/InAs/GaSb/AlAs structure. The cases without disorder (upper panel), with interface roughness (middle panel) and with Si-dopants (lower panel) are shown. 50 random realizations of disorder have been averaged. }
\label{fig6}
\end{figure}

The in-plane localization result from the strong admixture of the states by the disorder. In absence of scatterer, each state $\Psi_\nu$ is associated with a single plane wave of wavevector $\bm{k}$, but in disordered samples $\Psi_\nu$  contains several $k$ contributions in its expansion (\ref{eq_eigenstate}). In the previous section, we have shown the consequence of this adxmiture on the in-plane carrier motion, but this mixing also translates into a strong admixture of  the wavefunctions $\bm{\chi}_n(\bm{k},z)$ for the motion along the $z$ direction. To illustrate this effect, we define the normalized probability density of the eigenstates in each layer as:
\begin{equation}
  \zeta^{(\nu)}_\mathrm{InAs/GaSb}=\frac{\int_\mathrm{InAs/GaSb}|\tilde{\bm{\chi}}_\nu(z)|^2 dz}{\int_{-\infty}^{+\infty}|\tilde{\bm{\chi}}_\nu(z)|^2 dz}
  \label{eq_zeta}
\end{equation}
where $\tilde{\bm{\chi}}_\nu(z)$ is the $z$-part of the disordered wavefunctions obtained by integrating  $\Psi_\nu(\bm{\rho},z)$ over the $(x,y)$ plane. Figure~\ref{fig6} displays the energy dependence of $\zeta^{(\nu)}_\mathrm{InAs}$ and $\zeta^{(\nu)}_\mathrm{GaSb}$ for the spin-up orientation (similar results are obtained for the spin-down states, not shown here). One particularity of the type II alignment is that the confinement along the growth axis of the conduction and valence states does not occur in the same layer.
As shown in the upper panel of Fig.~\ref{fig6}, without disorder and far from the subband edges (for energies $E>135$~meV and $E<90$~meV), the conduction carriers are confined in the InAs layer while the valence carriers are confined in the GaSb layer.
Near the subband edges, or in other words in the inverted region ($90$~meV$<E<135$~meV), the confining layer varies from InAs to GaSb reflecting the band inversion and the hybridization of the valence and conduction states in this energy range. Values of $\zeta^{(\nu)}_\mathrm{InAs/GaSb}$ lower than $1$ and close to 0.5 indicate a delocalization of strongly hybridized wavefunctions over the two layers. As shown in Fig.~\ref{fig6}, the static disorder strengthens this intrinsic hybridization. The number of delocalized states increases and the hybridization region becomes wider. These features are again found to be weaker for interface roughness than for Si-dopants. 
Finally, it is important to remark that such delocalization effects along the growth axis are specific to inverted QWs. In non-inverted semiconductor QWs, where there is no electron-hole admixture, no disorder-induced hybridization is expected.

\section{Conclusion}
\label{sec6}

We have theoretically investigated the effects of the interface roughness and Si-dopants on the bulk states of an inverted InAs/GaSb QW. The combination of $\bm{k}\cdot \bm{p}$ calculations and numerical diagonalizations enable to get a deep understanding and a full characterization of the disordered energy spectrum and eigenfunctions of the system.
We found that an intentional doping leads to the opening of a mobility gap in the vicinity of the conduction and valence subband edges despite the closing of the energy gap due to the formation of in-gap bound states. This result is in agreement with the features found by the tight-binding approach and can be identified as being the origin of a decrease of the bulk conductance in actual samples.
The interface roughness, for its part, generates a potential that is too weak to significantly localize the bulk states and only perturbs marginally their energy spectrum. On the other hand, both types of scatterer affect the remaining extended states and strengthen the intrinsic hybridization of electrons and holes.

\begin{acknowledgements}
C-NB thanks Dr. K. Ohtani, Dr. A. Soluyanov and Pr. G. Bastard for fruitful and valuable discussions and gratefully acknowledges the support from the ERC under the project MUSiC. 
\end{acknowledgements}


\begin{thebibliography}{24}
\bibitem [{foo()}]{footnote}%
  \BibitemOpen
  \href@noop {} {\bibinfo  {journal} {An unperturbed state would be perfectly
  extended in the layer plane and would have a localization length equal to the
  sample size, in the present case 160 nm}\ }\BibitemShut {NoStop}%
\end{thebibliography}

%

\end{document}